\documentclass[12pt]{article}
\usepackage{epsf}
\def\beq{\begin{equation}}
\def\eeq{\end{equation}}
\def\ap#1#2#3 {Ann. Phys. (NY) {\bf#1} (19#2) #3}
\def\err#1#2#3 {{\it Erratum} {\bf#1} (19#2) #3}
\def\ib#1#2#3 {{\it ibid.} {\bf#1} (19#2) #3}
\def\ijmp#1#2#3 {Int. J. Mod. Phys. {\bf#1} (19#2) #3}
\def\jetp#1#2#3 {JETP Lett. {\bf#1} (19#2) #3}
\def\mpl#1#2#3 {Mod. Phys. Lett. {\bf#1} (19#2) #3}
\def\np#1#2#3 {Nucl. Phys. {\bf#1} (19#2) #3}
\def\pl#1#2#3 {Phys. Lett. {\bf#1} (19#2) #3}
\def\prep#1#2#3 {Phys. Rep. {\bf#1} (19#2) #3}
\def\prev#1#2#3 {Phys. Rev. {\bf#1} (19#2) #3}
\def\prl#1#2#3 {Phys. Rev. Lett. {\bf#1} (19#2) #3}
\def\sjnp#1#2#3 {Sov. J. Nucl. Phys. {\bf#1} (19#2) #3}
\def\spj#1#2#3 {Sov. Phys. JETP {\bf#1} (19#2) #3}
\def\spu#1#2#3 {Sov. Phys. Usp. {\bf#1} (19#2) #3}
\def\zp#1#2#3 {Zeit. Phys. {\bf#1} (19#2) #3}

\def\lsim{\raisebox{-0.6ex}{$\stackrel{\textstyle <}{\sim}$}}

\begin{document}
\begin{titlepage}
\begin{center}
{\Large \bf William I. Fine Theoretical Physics Institute \\
University of Minnesota \\}
\end{center}
\vspace{0.2in}
\begin{flushright}
FTPI-MINN-05/42-T \\
UMN-TH-2415-05 \\
September 2005 \\
\end{flushright}
\vspace{0.3in}
\begin{center}
{\Large \bf  $X(3872)$ diagnostics with  decays to $D {\bar D} \gamma$
\\}
\vspace{0.2in}
{\bf M.B. Voloshin  \\ }
William I. Fine Theoretical Physics Institute, University of
Minnesota,\\ Minneapolis, MN 55455 \\
and \\
Institute of Theoretical and Experimental Physics, Moscow, 117259
\\[0.2in]
\end{center}

\begin{abstract}
It is pointed out that in the decays $X(3872) \to D {\bar D} \gamma$ the
possible `molecular' component of the $X(3872)$ should give rise
essentially only to a $D^0 {\bar D}^0 \gamma$ final state with a
predictable spectrum, and should yield virtually no contribution to
decays into $D^+D^- \gamma$. The latter final state with charged $D$
mesons should however arise from the radiative decays of the
short-distance core of the $X(3872)$ through the transition $X(3872) \to
\psi(3770) \, \gamma$. Thus an observation of the radiative
decays of $X(3872)$ to pairs of neutral and charged $D$ mesons would provide an
insight into the internal dynamics of the $X(3872)$ resonance.
\end{abstract}

\end{titlepage}

\section{Introduction}
The narrow resonance $X(3872)$ observed through its decay channel $X \to \pi^+
\, \pi^- \, J/\psi$ in the $B \to X \, K$ decays\cite{belle1,babar1} and in
inclusive production in $p {\bar p}$ annihilation\cite{cdf,d0} attracts a great
theoretical and experimental interest. The peculiarity of this resonance is that
its mass is within approximately $0.6 \pm 1.0\,$MeV from the $D^0 {\bar D}^{*0}$
threshold, which strongly suggests that its internal composition can be
significantly contributed by a `molecular' state made of the charmed mesons. An
existence of such states of heavy hadrons was suggested\cite{ov} and
discussed\cite{drgg,nat,ek} long ago, and this idea was revived by the
observation of $X(3872)$\cite{cp,ps,mv,nat2}. The interpretation of $X(3872)$ as
largely a `molecular' state is further boosted by the observation\cite{belle2}
of the decay $X \to \pi^+ \, \pi^- \, \pi^0 \, J/\psi$, with a rate similar to
that of the discovery mode $X \to \pi^+ \, \pi^- \, J/\psi$. The co-existence of
these two decay modes implies that the isospin is badly broken in the resonance
$X(3872)$, which would be impossible if it were just another charmonium state.
Furthermore, the observation\cite{belle2} of the decay $X \to \gamma \, J/\psi$
implies positive C parity of $X$, and a detailed study\cite{belle3} of  the
decay $X(3872) \to \pi^+ \, \pi^- \, J/\psi$ most strongly  favors the
identification of its quantum numbers as $J^{PC}=1^{++}$. Such quantum numbers
as well as a substantial isospin violation agree well\cite{mv2} with that a
molecular C-even S-wave state of neutral charmed mesons: $D^0 {\bar D}^{*0} +
D^{*0} {\bar D}^0$ makes up a significant part of the wave function of the
$X(3872)$ resonance.

There is however no reason to expect that there is only the molecular
component in the wave function that determines all of the properties of
the $X(3872)$ boson. Rather one should consider the wave function in
terms of a general Fock decomposition:
\beq
\psi_X=a_0 \, \psi_0 + \sum_i a_i \, \psi_i~,
\label{fock}
\eeq
where $\psi_0$ is the state of the neutral $D$ mesons $(D^0 {\bar
D}^{*0}+ {\bar D^0}  D^{*0})/\sqrt{2}$, while $\psi_i$ refer to `other'
hadronic states. Due to the extreme proximity of the mass of $X$ to the
$D^0 {\bar D}^{*0}$ threshold, the $\psi_0$ part should be dominant at
long distances. Indeed, assuming that the mass of $X$ is below the
threshold by the binding energy $w$: $m_{D^0}+m_{D^{*0}}-M_X=w$, the
spatial extent of the $\psi_0$ is determined as $(m_D \, w)^{-1/2}
\approx 5 \, {\rm fm} \, (1 \, {\rm MeV}/  w)^{1/2} $, and $\psi_0$ thus
describes the `peripheral' part of the wave function, in fact beyond the
range of strong interaction. On the other hand, the `other' states in
the sum in the Fock decomposition (\ref{fock}) are localized at shorter
distances and constitute the `core' of the $X(3872)$ wave function. In
other terms one may think of this picture as that of a mixing in
$X(3872)$ of the molecular component $D^0 {\bar D}^{*0} + D^{*0} {\bar
D}^0$ with `other' states, such as e.g. a `pure' $c {\bar c}$
charmonium, which then has to be in a $^3P_1$ state, also favored by the
heavy quark spin selection rule\cite{mv2}.

Since the internal composition of the $X(3872)$ can be quite different
at different distances, one or the other part of the Fock decomposition
(\ref{fock}) may be important in specific processes. It appears that the
pionic transitions from $X(3872)$ to $J/\psi$ are determined by a long
distance dynamics, where the $D^0 {\bar D}^{*0} + D^{*0} {\bar D}^0$
component dominates, so that the isospin states are mixed, and the
$\pi^+ \pi^-$ and $\pi^+ \pi^- \pi^0$ transitions have approximately the
same strength. The production of $X$ however is determined by short
distances, and proceeds through the core component\cite{mv2}, which is
approximately an isospin singlet, as evidenced\cite{babar2} by a
comparable relative rate of the decays $B^+ \to X \, K^+$ and $B^0 \to X
\, K^0$, while contribution of the molecular $D^0 {\bar D}^{*0} + D^{*0}
{\bar D}^0$ would correspond to a strong suppression\cite{bk,suzuki} of
the  $B^0 \to X \, K^0$ decay in comparison with $B^+ \to X \, K^+$.

The purpose of the present paper is to point out that both the
peripheral molecular component of the $X(3872)$ and, to some extent, its
core can be studied in the decays $X(3872) \to D {\bar D} \gamma$. It
has been previously noted\cite{mv} that the rates of the decays $X \to
D^0 {\bar D}^0 \pi^0$ and $X \to D^0 {\bar D}^0 \gamma$ are sensitive to
the coordinate wave function of the molecular component due to the
interference between the decays of $D^{*0}$ and ${\bar D}^{*0}$ mesons
in a state of definite C parity. Here the decays of $X$ into $D^0 {\bar
D}^0 \gamma$ as well as into $D^+ D^- \gamma$ are considered in more
detail, including the effects of re-scattering $D^0 {\bar D}^0
\leftrightarrow D^+ D^-$ and of the charmonium resonance $\psi(3770)$
strongly coupled to the $D {\bar D}$ channel. As a result it will be
argued here that at the photon energy $\omega$ near that in the decay
$D^{*0} \to D^0 \gamma$, $\omega_0 \approx 137 \,$MeV, the contribution
of the molecular component of $X(3872)$ should dominate with the
radiative decay going almost exclusively into the channel with neutral
mesons: $D^0 {\bar D}^0 \gamma$, and the underlying process being the
decays $D^{*0} \to D^0 \gamma$ and ${\bar D}^{*0} \to {\bar D}^0
\gamma$. At a somewhat lower photon energy $\omega \approx 100\,$MeV the
final state with charged mesons, $D^+ D^- \gamma$, should also appear
due to the underlying process of radiative transition from a charmonium
`core' component of $X(3872)$ to $\psi(3770) \gamma$. Thus a measurement
of the photon spectra in the radiative decays of $X$ into $D^0 {\bar
D}^0 \gamma$ and $D^+ D^- \gamma$ would provide an information on both
the peripheral molecular component and the `core' of the $X(3872)$.

The rest of the paper is organized as follows. In Sec.2 the wave function of the
$D {\bar D}^* + D^* {\bar D}$ mesons within the $X(3872)$ resonance is discussed
as well as the possible states making up the `core'. In Sec.3 the contribution
of the peripheral meson component to the decay $X \to D^0 {\bar D}^0 \gamma$ is
calculated, and in Sec.4 it is argued that the peripheral contribution to the
decay $X \to D^+D^- \gamma$ is very small, both due to the direct decays
$D^{*\pm} \to D^\pm \gamma$ as well as due to the rescattering $D^0 {\bar D}^0
\to D^+ D^-$. In Sec.5 the radiative transition $X(3872) \to \psi(3770) \gamma$
as proceeding due to a charmonium `core' of the $X$ resonance is considered,
which results in the decay $X \to D {\bar D} \gamma$ with both the neutral and
the charged mesons. Finally, Sec.6 contains a discussion and concluding remarks.

\section{States in the Fock decomposition of $X(3872)$}
The extreme proximity of the mass of $X(3872)$ to the $D^{*0} {\bar D}^{*0}$
guarantees that the mesons in the $D^0 {\bar D}^{*0} + D^{*0}
{\bar D}^0$ component move freely beyond the range of the strong interaction,
where their wave function in the coordinate space is given by
\beq
\phi_n(r)=c \, {\exp (-\kappa_n \, r) \over r}~,
\label{phic}
\eeq
where $\kappa_n$ is determined by the binding energy $w$ and the reduced mass
$m_r \approx 966\,$MeV in the $D^0 {\bar D}^{*0}$ system as $\kappa_n = \sqrt{2
\, m_r \, w}$. The normalization coefficient $c$ determines the statistical
weight of the $D^0 {\bar D}^{*0} + D^{*0}
{\bar D}^0$ component in $X(3872)$, and its definition is correlated with that
of the coefficient $a_0$ in the Fock decomposition (\ref{fock}). We resolve this
ambiguity in the definition by requiring that the coordinate wave function of
the neutral meson pair be normalized to one, so that the statistical weight of
the state $(D^0 {\bar D}^{*0} + D^{*0}
{\bar D}^0)/\sqrt{2}$ is given as $|a_0|^2$. If the wave function of the form
(\ref{phic}) is used down to $r=0$, this requirement corresponds to
$c=\sqrt{\kappa_n/(2\pi)}$.

The dominance of the $D^0 {\bar D}^{*0} + D^{*0}{\bar D}^0$ at long distances
translates into a substantial isospin violation in the processes determined by
the `peripheral' dynamics, examples of which are apparently the observed decays
$X \to \pi^+  \pi^- J/\psi$ and  $X \to \pi^+  \pi^- \pi^0 J/\psi$. It is quite
likely however that this isospin-breaking behavior is only a result of the
`accidentally' large mass difference $\delta \approx 8 \,$MeV between $D^+
D^{*-}$ and $D^{*0} {\bar D}^{*0}$. Therefore
it is natural to expect that at shorter distances within the range of the strong
interaction the isospin symmetry is restored and at those distances the wave
function of $X(3872)$ is dominated by $I=0$. In this picture the wave function
of a $D^+ D^{*-}+D^-D^{*+}$ state within the region beyond the range of the
strong interaction can be found from eq.(\ref{phic}) by requiring that at short
distances the pairs of charged and neutral mesons combine into an $I=0$
state\footnote{The effects of the Coulomb interaction between the charged mesons
are neglected here.}, so that the wave function of the charged meson pair has
the form
\beq
\phi_c(r)= c \, {\exp (-{\kappa_c} \, r) \over r}~,
\label{phicc}
\eeq
where ${\kappa_c} = \sqrt{2 m_r \, (\delta+w)} \approx 125 \,$MeV. It should be
noticed that both the neutral (eq.(\ref{phic})) and the charged
(eq.(\ref{phicc})) meson wave function have the same normalization factor $c$
(determined by $\kappa_n$) and differ only in the exponential power.
This simple picture allows one to estimate the relative statistical weight of
the charged and neutral $D$ meson components in the $X(3872)$:
\beq
\lambda \equiv {\left |\langle X | D^+ D^{*-}+D^-D^{*+}\rangle \right |^2 \over
\left | \langle X | D^0 {\bar D}^{*0} + D^{*0}{\bar D}^0 \rangle \right |^2} =
{\kappa_n \over {\kappa_c}}~.
\label{wrat}
\eeq

Clearly, the wave functions in eq.(\ref{phic}) and eq.(\ref{phicc}) cannot be
applied at short distances in the region of strong interaction, where the mesons
overlap with each other and cannot be considered as individual particles. In
order to take into account this behavior an `ultraviolet' cutoff should be
introduced. One widely used method for introducing such cutoff is to consider
the meson wave functions only down to a finite distance $r_0$, at which distance
the boundary condition of the state being that with $I=0$ is imposed (a
discussion of a similar situation can be found e.g. in Ref.\cite{mv3}). An
alternative, somewhat more gradual cutoff, described by parameter $\Lambda$, can
be introduced\cite{suzuki} by subtracting from the wave functions (\ref{phic})
and (\ref{phicc}) an expression $c \, e^{-\Lambda r}/r$. It should be noticed,
that such regularization also results in a modification of the normalization
coefficient $c$, which for the gradual cutoff takes the form
\beq
c=\sqrt{\kappa_n \over 2 \pi} \, {\sqrt{\Lambda \, (\Lambda + \kappa_n)} \over
\Lambda - \kappa_n}~.
\label{norm}
\eeq
One can also readily see, that an introduction of any such cutoff eliminates
relatively more of the charged meson wave function than of the neutral one, thus
reducing the estimate of the relative statistical weight as compared to that in
eq.(\ref{wrat}), so that eq.(\ref{wrat}) gives in fact the upper bound for the
ratio.

The structure of the `core' of $X(3872)$ in the region of strong interaction can
be quite complicated. As mentioned above, it is natural to expect that the
states in the `core' are dominantly isotopic scalars, which however still leaves
the possibility that the essential states contain light-quark pairs and/or
gluons in addition to the $c {\bar c}$ charm quark pair. However the simplest
configuration with just a charmonium $^3 P_1$ state is also allowed by the
quantum numbers, and in what follows only the contribution of the charmonium
`core' to the discussed properties is considered. In other words, the states in
the Fock decomposition (\ref{fock}) taken into account in this paper are the
charm meson pairs at the `peripheral' distances and a $J^{PC}=1^{++}$ charmonium
$c {\bar c}$ pair in the `core'. It can be mentioned that such simplified
structure corresponds to the approach\cite{elq} based on channel mixing between
the charmed meson pairs and the states of charmonium.

\section{Peripheral contribution to the decay  $X(3872) \to D {\bar D}
\gamma$}
The dominant peripheral contribution to the decay $X(3872) \to D^0 {\bar D}^0
\gamma$ arises from the radiative decays of individual vector mesons:
 $D^{*0} \to D^0 \gamma$ and ${\bar D}^{*0}
\to {\bar D}^0 \gamma$\cite{mv}, while the peripheral contribution to the decay
$X(3872) \to D^+ D^- \gamma$ originates dominantly from the decays $D^{*\pm} \to
D^\pm \gamma$. As will be argued, the rate for the latter process should be
expected considerably smaller than for the radiative decay with neutral mesons.
It will be also argued that the effects of a rescattering $D^0 {\bar D}^0
\leftrightarrow D^+ D^-$ should be very small in the kinematical region, where
the peripheral contribution dominates. In this section the rescattering effects
are completely neglected.

 The amplitude of the radiative decay can be written as
\beq
A(D^{*} \to D \gamma)=\mu \, \epsilon_{ijk} \,\varepsilon_i  \, k_j
\, a_k~,
\label{ad0}
\eeq
where ${\vec a}$ and ${\vec \varepsilon}$ are the polarization
amplitudes for the photon and the vector meson, ${\vec k}$ is the
momentum of the photon, and the non-relativistic normalization for the
heavy-meson states is assumed, so that the decay rate is expressed in
terms of the parameter $\mu$ as
\beq
\Gamma(D^* \to D \gamma) = {|\mu|^2 \, \omega_0^3
\over 3 \pi}~.
\label{gd0}
\eeq
The widths of the decays $D^{*0} \to D^0 \gamma$ and $D^{*\pm} \to D^\pm \gamma$
can be readily deduced from the data\cite{pdg}: $\Gamma(D^{*\pm} \to D^\pm
\gamma) = 1.5 \pm 0.5\,$KeV, $\Gamma(D^{*0} \to D^0 \gamma)=26 \pm 6\,$KeV. One
can see from these numbers that the transition magnetic moment $\mu$ for the
charged meson decay is substantially smaller than that for the neutral meson
radiative decay: $|\mu_c|^2 \ll |\mu_n|^2$.

If the final-state interaction between the pseudoscalar mesons is
neglected, the amplitude of the decay $X(3872) \to D {\bar D} \gamma$ due to the
peripheral
$D {\bar D}^{*} + D^{*} {\bar D}$ component can be written
as\cite{mv}
\beq
A_0(X \to D {\bar D} \gamma)={\mu \, a_0  \over \sqrt{2}} \,
\epsilon_{ijk} \,\varepsilon_i  \, k_j \, a_k \, \left [ \phi \left
({{\vec k} \over 2} + {\vec p} \right )- \phi\left ({{\vec k} \over 2} -
{\vec p}\right ) \right ]~.
\label{a0}
\eeq
Here $a_0$ is the weight amplitude in eq.(\ref{fock}) for the
`molecular' state in $X(3872)$, ${\vec p}=({\vec p}_D-{\vec p}_{\bar
D})/2$ is the momentum of the $D$ meson in the c.m. frame of the $D
{\bar D}$ system,  and $\phi({\vec q})$ is the
wave function of the peripheral meson component in the momentum space. Clearly,
one should use the neutral-meson wave function $\phi_n$ and the corresponding
transition magnetic moment $\mu_n$ for the decay $X \to D^0 {\bar D}^0 \gamma$
and the charged-meson one wave function $\phi_c$ and $\mu_c$ for the decay $X
\to D^+ D^- \gamma$.
The two terms in the square braces in
eq.(\ref{a0}) correspond respectively to the processes $D^{*} \to D
\gamma$ and ${\bar D}^{*} \to {\bar D} \gamma$, and the relative
minus sign is due to the opposite C parity of the photon and of the
initial state of the mesons.

The differential decay rate found from eq.(\ref{a0}) can be written in
terms of the radiative width of $D^{*}$ meson (\ref{gd0}) as
\beq
{\rm d}\Gamma(X \to D {\bar D} \gamma)=\Gamma(D^* \to D \gamma) \, {a_0^2 \over
2}
\, \left ( {\omega \over \omega_0} \right )^3 \, \left [ \phi \left
({{\vec k} \over 2} + {\vec p} \right )- \phi\left ({{\vec k} \over 2} -
{\vec p}\right ) \right ]^2 {{\rm d}^3 p \over (2 \pi)^3}~,
\label{dgam0}
\eeq
where $\omega_0$ stands for the photon energy in the corresponding decay $D^*
\to D \gamma$ ($\omega_0=137\,$MeV in the decay $D^{*0} \to D^0 \gamma$ and
$\omega_0=136\,$MeV in the radiative decay of $D^{*\pm}$), the photon energy
$\omega$ and the momentum ${\vec p}$ in eq.(\ref{dgam0}) are related
by the energy conservation relation
\beq
\omega + {\omega^2 \over 4 \, m_D}+{{\vec p}\,^2 \over m_D}= \Delta
\label{cons}
\eeq
with
$\Delta=M_X-2 \, m_D$.

The photon spectrum given by eq.(\ref{dgam0}) will be further  discussed
in some detail. It is clear on general grounds that the dominant contribution
to the total rate comes from the region where $\omega$ is very close to
$\omega_0$. Such values of $\omega$ are kinematically allowed for the $X \to D^0
{\bar D}^0 \gamma$ decay and are forbidden for the decay $X \to D^+ D^- \gamma$.
Thus for the former decay the total rate can be
approximated by setting $\omega=\omega_0$ in eq.(\ref{dgam0}), which results in
\begin{eqnarray}
\Gamma(X \to D^0 {\bar D}^0 \gamma)&=&\Gamma(D^{*0} \to D^0 \gamma) \, a_0^2
\left [ 1- \int
\phi_n \left ({{\vec k} \over 2} + {\vec p} \right ) \phi_n\left ({{\vec k}
\over 2} - {\vec p}\right ) \, {{\rm d}^3 p \over (2 \pi)^3} \right ]
\nonumber \\
&=&\Gamma(D^{*0} \to D^0 \gamma) \, a_0^2 \left [ 1- 4\pi \,\int \phi_n^2(r)\,
{\sin kr
\over kr}  \, r^2 \, {\rm d} r \right ]~.
\label{gtn}
\end{eqnarray}
In this formula the normalization condition and the absence of an angular
dependence of an $S$ state wave function are explicitly taken into account.

With the unregularized free-motion wave functions (\ref{phic}) and (\ref{phicc})
one readily finds the contribution of the peripheral meson component to the
total rate of the discussed decays as\cite{mv}\footnote{This equation corrects
the corresponding formula of Ref.\cite{mv}, which erroneously contains an extra
factor of two.}
\beq
\Gamma(X \to D^0 {\bar D}^0 \gamma)=\Gamma(D^{*0} \to D^0 \gamma) \, a_0^2 \left
[ 1-{2 \kappa_n \over \omega_0} \, \arctan {\omega_0 \over 2 \kappa_n} \right
]~.
\label{gtot}
\eeq

The regularization at short distances in fact enhances the total rate of the
dominant decay $X \to D^0 {\bar D}^0 \gamma$ when the rate is expressed in terms
of $a_0^2$. Indeed, the integral weight in the negative term in the interference
factor in eq.(\ref{gtn}) is the largest at short distances. Thus, eliminating by
regularization a short-distance part of the meson wave function reduces this
negative term and enhances the rate. In particular for the gradual cutoff
regularization the expression for the total decay rate takes the form
\begin{eqnarray}
&&\Gamma(X \to D^0 {\bar D}^0 \gamma) =
\Gamma(D^{*0} \to D^0 \gamma) \, a_0^2 \times  \nonumber \\
&&\left [ 1-
 {2 \kappa_n \over \omega_0} \, {\Lambda \, (\Lambda+\kappa_n) \over (\Lambda -
\kappa_n)^2} \left ( \arctan {\omega_0 \over 2 \kappa_n} + \arctan {\omega_0
\over 2 \Lambda} - 2 \arctan {\omega_0 \over \Lambda+\kappa_n} \right ) \right
]~.
\label{gtnr}
\end{eqnarray}
At $\omega_0=137\,$MeV and $\kappa_n \approx 44\,$MeV (corresponding to the
binding energy $w = 1\,$MeV) the numerical value of the expression in the square
braces, the interference factor, varies between 0.36 at $\Lambda \to \infty$ and
$0.61$ at $\Lambda=200\,$MeV.

\begin{figure}[ht]
\begin{center}
 \leavevmode
    \epsfxsize=12cm
    \epsfbox{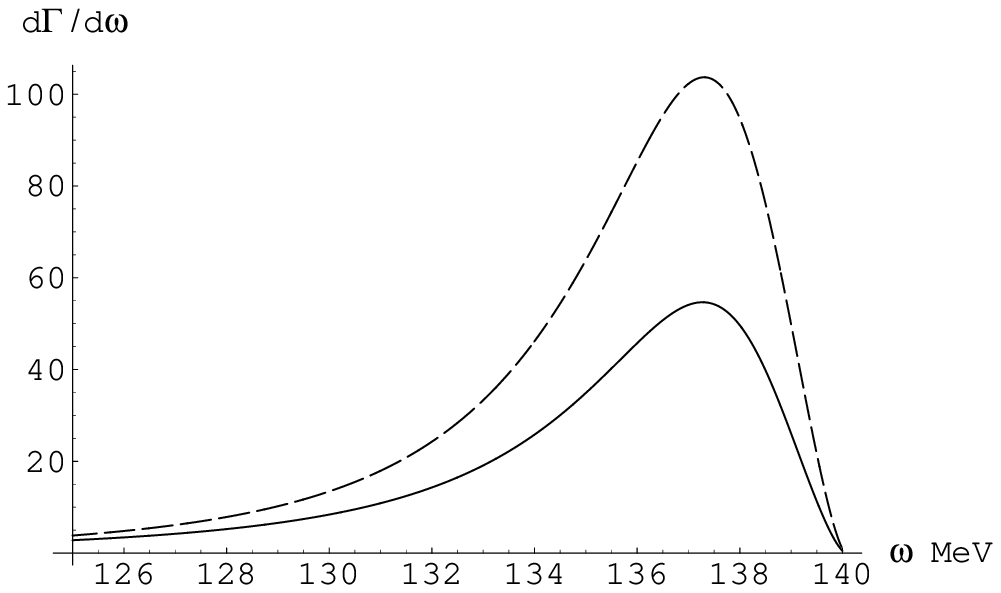}   
    \caption{The photon spectrum in the decay $X(3872) \to D^0 {\bar D}^0
\gamma$ for $\kappa_n = 44\,$MeV  at $\Lambda=200\,$MeV (dashed) and at $\Lambda
\to \infty$ (solid). The vertical scale is in arbitrary units.}
\end{center}
\end{figure}

The photon spectrum in the decay $X \to D^0 {\bar D}^0 \gamma$ can be found from
eq.(\ref{dgam0}) and eq.(\ref{cons}) using the regularized form of the wave
function in the momentum space,
\beq
\phi_c({\vec q})=4 \pi \, c \, \left ( {1 \over q^2 + \kappa_n^2}- {1 \over q^2
+ \Lambda^2} \right ) ~.
\label{phiq}
\eeq
The final expression is found after approximating the solution to
eq.(\ref{cons}) as $p \approx \sqrt{m_D (\omega_1-\omega)}$ with $\omega_1
\approx 140\,$MeV being the maximal kinematically allowed photon energy in the
decay and after performing the integration over the angle between the photon
momentum and ${\vec p}$. At finite regularization parameter $\Lambda$ the
resulting expression is quite lengthy. However the effect of a finite $\Lambda$
in the essential part of the spectrum, i.e. near $\omega=\omega_0$ to a very
good accuracy reduces to an overall rescaling, according to the previously
mentioned enhancement of the decay rate for a regularized wave function. This
behavior is illustrated in Fig.1.

In the limit of no regularization, i.e. at $\Lambda \to \infty$, the expression
for the spectrum takes a more concise form:
\beq
{{\rm d} \Gamma \over {\rm d} \omega} (X \to D^0 {\bar D}^0 \gamma)=
\Gamma(D^{*0} \to D^0 \gamma) \, a_0^2 \, {\omega^3 \over \omega_0^3 } \, {2 \,
m_D \, \kappa_n \over \pi} \, p \, f(p^2 + \omega^2/4 +\kappa_n^2, p \,
\omega)~,
\label{spectr}
\eeq
where
\beq
f(x,y)={1 \over x^2-y^2}- {1 \over x \, y} \, {\rm arctanh} {y \over x}~,
\label{funcf}
\eeq
and the momentum $p$  is related to $\omega$ by the energy conservation relation
(\ref{cons}). These formulas allow to illustrate the dependence of the spectrum
on the mass gap $w$ between the $X(3872)$ and the $D^0 {\bar D}^{*0}$ threshold,
as shown in Fig.2 for two representative values of $w$: $w \approx 1\,$MeV
($\kappa_n=44$) and $w \approx 0.3\,$MeV ($\kappa_n=24\,$MeV). One can readily
see the expected behavior: with decreasing $w$ the spectrum becomes more
narrowly peaked around $\omega_0$ and is enhanced due to diminishing destructive
interference.
\begin{figure}[ht]
\begin{center}
 \leavevmode
    \epsfxsize=12cm
    \epsfbox{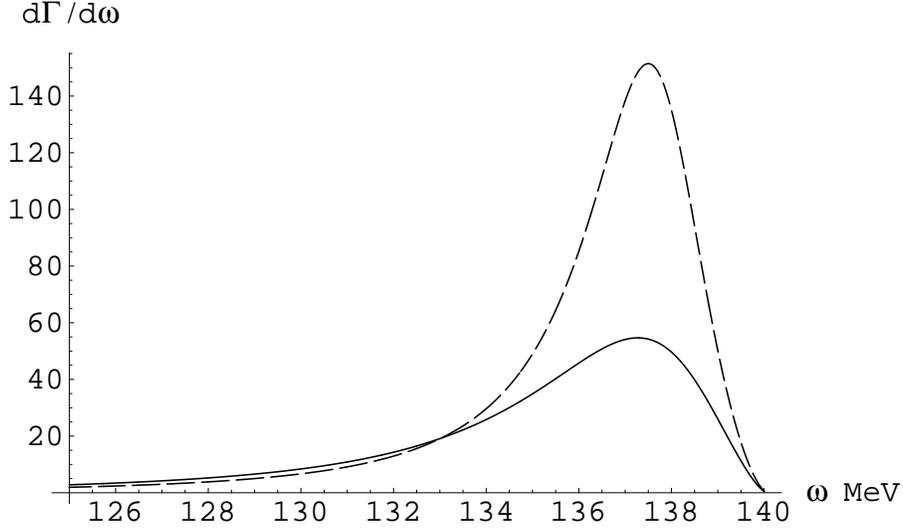}   
    \caption{The photon spectrum in the decay $X(3872) \to D^0 {\bar D}^0
\gamma$ for $\kappa_n = 44\,$MeV (solid) and for $\kappa_n=24\,$MeV (dashed),
both at $\Lambda \to \infty$. The vertical scale is in arbitrary units.}
\end{center}
\end{figure}

\section{Peripheral contribution to the decay $X \to D^+ D^- \gamma$ and the
effects of final-state interaction}
For the decay $X \to D^+ D^- \gamma$ the maximal allowed energy of the photon is
$\omega_2 \approx 131\,$MeV, which is below the photon energy $\omega_0 \approx
136\,$MeV in the decay $D^{*\pm} \to D^\pm \gamma$, so that the peak in the
spectrum, corresponding to a free-meson decay is kinematically inaccessible.
Thus the peripheral contribution to this decay has to be calculated by
integrating the differential decay rate. The latter rate can be estimated from a
formula similar to eq.(\ref{spectr}):
\beq
{{\rm d} \Gamma \over {\rm d} \omega} (X \to D^+ D^- \gamma)= \Gamma(D^{*\pm}
\to D^\pm \gamma) \, a_0^2 \, \lambda \, {\omega^3 \over \omega_0^3 } \, {2 \,
m_D \, \kappa_c \over \pi} \, p \, f(p^2 + \omega^2/4 +\kappa_c^2, p \,
\omega)~,
\label{spc}
\eeq
and the total rate found from this expression is very small, corresponding to
the bound (on the peripheral contribution to the rate) at all values of $w$
below 1\,MeV:
\beq
\left . {\Gamma(X \to D^+ D^- \gamma) \over \Gamma(X \to D^0 {\bar D}^0 \gamma)
} \right |_{\rm peripheral} < 10^{-3}~.
\label{prat}
\eeq
Such strong suppression of the decay $X \to D^+ D^- \gamma$ is a result of three
suppression factors: the small rate of $D^{*\pm} \to D^\pm \gamma$, the small
statistical weight of the charged meson pair in the $X(3872)$ relative to that
of the neutral pair (eq.(\ref{wrat})), and the stronger destructive interference
in emission of photon by the charged meson pair in the $X(3872)$, the latter
suppression factor being combined with the discussed kinematical restriction on
the photon energy, which further suppresses the decay rate.

The effects of the rescattering of the $D$ mesons in the final state were
completely neglected in the previous estimates. It can be argued that these
effects could give rise only to a very small rate of the decay $X \to D^+ D^-
\gamma$ and have very little impact on the photon spectrum in the decay $X \to
D^0 {\bar D}^0 \gamma$ arising from the peripheral component of the $X(3872)$.

Indeed, the $D$ meson pair produced in the discussed decay is $C$ odd and
therefore the relative angular momentum of the meson is also odd. It can be
noted that the decay rate is dominated by the kinematical region, where the
relative momentum ${\vec p}$ of the $D$ mesons is small, so that the main
contribution to the rate comes from the states where the $D$ meson pair is in
the lowest partial wave with odd angular momentum, i.e. in the $P$ wave. One can
readily verify that in the previous estimates of the peripheral contribution to
the rate of the decay $X \to D^0 {\bar D}^0 \gamma$ at $w < 1\,$MeV more than
93\% of the rate is associated with the production of the meson pair in the $P$
wave. In the kinematical region of this decay with the photon energy $\omega$
above the maximal energy $\omega_2$ for the decay $X \to D^+ D^- \gamma$ only
the elastic scattering between $D^0$ and  ${\bar D}^0$ is possible, so that only
the phases of the partial-wave production amplitudes can be modified by small
scattering phases, but not the differential or the total rate of the decay. At
$\omega < \omega_2$ the $D$ meson pair is above the threshold for $D^+ D^-$ and
the charged meson pair can in principle be produced by the rescattering $D^0
{\bar D}^0 \to D^+ D^-$. However in this region the amplitude of the decay $X
\to D^0 {\bar D}^0 \gamma$ is already small, and the rescattering effect is
further suppressed by the small $P$-wave factor $p_\pm^3$ with $p_\pm$ being the
relative momentum of the $D^+ D^-$ pair. The rescattering of the $D$ mesons is
kinematically suppressed near the threshold below the $\psi(3770)$ resonance,
and its effect can be rather conservatively estimated as resulting in the rate
of the charged meson pair production $\Gamma(X \to D^+ D^- \gamma)$ being less
than about 1\% of $\Gamma(X \to D^0 {\bar D}^0 \gamma)$. Although this bound is
considerably larger than the estimate in eq.(\ref{prat}) without the
rescattering, it is still quite small in absolute terms.  The mixing between the
pairs of neutral and charged $D$ mesons becomes of order one in the $\psi(3770)$
resonance, which corresponds to $\omega \approx 100\,$MeV. However in this
kinematical region the estimated in eq.(\ref{spectr}) peripheral contribution is
already very small.

\section{The `core' decay $X(3872) \to \psi(3770) \gamma$}

At the $\psi(3770)$ resonance the c.m. momentum of the $D$ mesons is quite
large: $p_\pm \approx 242\,$MeV and $p_0 \approx 275\,$MeV, so that not only the
estimate of the peripheral contribution to the decay $X \to D {\bar D} \gamma$
is small, but it also becomes rather unreliable, since the amplitude in this
kinematical region is sensitive to the short-distance `core' of the $X(3872)$.
In the picture discussed above, where this `core' is a $^3 P_1$ charmonium, it
is natural to describe the radiative decays as due a charmonium radiative
transition $^3P_1 \to \psi(3770) \gamma$ with subsequent decay $\psi(3770) \to D
{\bar D}$. Clearly, this mechanism should give the relative yield of the charged
and neutral pairs as is measured in the decay of the $\psi(3770)$
resonance\cite{cleo1}:
\beq
\left . {\Gamma(X \to D^+ D^- \gamma) \over \Gamma(X \to D^0 {\bar D}^0 \gamma)
} \right |_{\rm core}={\Gamma[\psi(3770) \to D^+  D^-]\over \Gamma[\psi(3770)
\to D^0 {\bar D}^0]}=0.776 \pm
0.024^{+0.014}_{-0.006}~.
\label{datb}
\eeq

The estimate of the absolute decay rate provided by this mechanism obviously
depends on two factors: the amplitude of the transition $^3P_1 \to \psi(3770)
\gamma$ and the statistical weight $|a_{c \bar c}|^2$ of the $^3P_1$ charmonium
in the wave function of the $X(3872)$ resonance. Needless to mention that any
estimate of each of these factors is necessarily quite approximate. Perhaps the
best way of estimating the amplitude of the radiative transition $^3P_1 \to
\psi(3770) \gamma$ is to use the recent data\cite{cleo2} on a similar transition
$\psi(3770) \to \chi_{c1} \gamma$:
\beq
\Gamma[\psi(3770) \to \chi_{c1} \gamma]=75 \pm 14 \pm 13\,{\rm KeV}~,
\label{datc}
\eeq
and to rescale the rate as $\omega^3$, assuming that the wave function of the
$^3P_1$ `core'  charmonium state in $X(3872)$ is similar to that in the
$\chi_{c1}$ resonance. Such estimate gives
\beq
\Gamma[X(3870) \to \psi(3770) \gamma] \approx |a_{c \bar c}|^2 \times 5 \, {\rm
KeV}~.
\label{gxp}
\eeq

In the approximation where the core of $X(3872)$ is a $^3P_1$ charmonium similar
to $\chi_{c1}$ the amplitude $a_{c \bar c}$ can be estimated from the rate of
production of the $X$ resonance in $B$ decays relative to $\chi_{c1}$, since in
$B$ decays $X(3872)$ is dominantly produced through its core\cite{mv2,suzuki},
rather than through the peripheral meson component. In particular, it is
known\cite{pdg} that ${\cal B}(B \to \chi_{c1} K) \approx {\cal B}(B \to \psi'
K)$, and also that\cite{belle1}
\beq
{{\cal B}(B^+ \to K^+ \, X) \, {\cal B}(X \to \pi^+ \, \pi^- \, J/\psi) \over
{\cal B}(B^+ \to K^+ \, \psi^{\prime}) \, {\cal B}(\psi^{\prime} \to \pi^+ \,
\pi^- \, J/\psi)} = 0.063 \pm 0.014~.
\label{dat}
\eeq
Therefore one can estimate
\beq
|a_{c \bar c}|^2 \approx 0.06  \, {{\cal B}(\psi^{\prime} \to \pi^+ \,
\pi^- \, J/\psi) \over {\cal B}(X \to \pi^+ \, \pi^- \, J/\psi)} \approx {0.02
\over {\cal B}(X \to \pi^+ \, \pi^- \, J/\psi)}~.
\label{acc}
\eeq
The recent experimental lower limit\cite{babar3} on the branching fraction, $
{\cal B}(X \to \pi^+ \, \pi^- \, J/\psi) > 4.2\%$, then
implies $|a_{c \bar c}|^2 \, \lsim \, 0.5$, with a more realistic value,
perhaps, being in the neighborhood of 0.1.

\section{Discussion}

The estimates presented here illustrate that a study of the decays $X(3872) \to
D {\bar D} \gamma$, if experimentally feasible, may provide important
information about the internal structure of the $X$ resonance. In particular, it
may be quite helpful in finding out the relative weight of the charmonium state
and the molecular one in the wave function describing the composition of
$X(3872)$. Recently the dilemma of whether this resonance is just another
charmonium level or a `molecule' was discussed in detail in Ref.\cite{suzuki}.
It is likely however that we will not have to resolve it in one way or the
other, and that the $X$ resonance is a part of both. In this picture the two
competing components contribute to different properties of the $X$: soft
processes such as the decays of the type $X \to \pi^+ \, \pi^- \, J/\psi$ and $X
\to \pi^+ \, \pi^- \, \pi^0 \, J/\psi$ proceed due to the peripheral molecular
part with a strong isospin violation, while the hard processes such as
production of $X(3872)$ in $B$ decays or in hadronic collisions are dominantly
due to the charmonium `core'. In this picture the $X$ resonance should behave as
an isoscalar in the hard processes. This implies in particular, that one should
expect\cite{mv2,mv4} approximately equal rates of the decays $B^+ \to X K^+$ and
$B^0 \to X K^0$, which expectation is in agreement with the most recent
data\cite{babar2}. Furthermore, if the core of $X(3872)$ is similar to the
$^3P_1$ charmonium state $\chi_{c1}$, this similarity can be tested in other
production processes where the yield of the $\chi_{c1}$ is known, e.g. in the
decays $B \to X K^*$ vs. the decays $B \to \chi_{c1} K^*$.

The decays $X \to D {\bar D} \gamma$ have the advantage that they receive
contribution from both the peripheral component and the core in different parts
of the photon spectrum. The peripheral contribution gives rise to essentially
only the decay with neutral mesons in the final state, $X \to D^0 {\bar D}^0
\gamma$, with the spectrum peaking near the photon energy $\omega_0 \approx
137\,$MeV as in the free $D^{*0}$ radiative decay. The charmonium core
contribution however is peaked near the photon energy $\omega \approx 100\,$MeV,
corresponding to the transition $X \to \psi(3770) \gamma$. In the latter peak
the ratio of the yield of the pairs of charged and neutral mesons is determined
by the decay properties of the $\psi(3770)$ resonance (eq.(\ref{datb})).  The
ratio of the rates under those peaks, if measured, would allow to quantitatively
estimate, using the equations (\ref{gtot}), (\ref{gtnr}), and (\ref{gxp}), the
statistical weights $|a_0|^2$ and $|a_{c \bar c}|^2$, while the shape of the
peak in the spectrum near $\omega_0$ may provide further information on the
details of the wave function of the `molecular' component.

\section*{Acknowledgments}
A part of this paper was done at the Aspen Center for Physics.

\noindent
This work is supported in part by the DOE grant DE-FG02-94ER40823.

\end{document}